\let\orilabel\label % Save the original LaTeX kernel definition - to fix hyperref: FEDE

\documentclass[%
10pt,
amsmath,
amssymb,
aps,
pra,
floatfix, % to fix floats
showkeys % if not - comment out
]{revtex4-2}
\let\label\orilabel % Restore the kernel definition over REVTEX's version - to fix hyperref: FEDE

\usepackage{graphicx} % Include figure files
\usepackage{amsfonts}
\usepackage{amssymb} % new - if errors remove
\usepackage{amsmath}
\usepackage{amsthm}
\usepackage[english]{babel}
\usepackage{booktabs}
\usepackage{braket}
\usepackage{enumerate}
\usepackage{graphicx}
\usepackage{xr-hyper}
\usepackage{hyperref} % Now hyperref/nameref will load cleanly
\usepackage{mathtools}
\usepackage{orcidlink}
\usepackage{physics}
\usepackage{dcolumn}% Align table columns on decimal point
\usepackage{bm}% bold math
\usepackage{cleveref}
\usepackage{graphicx}
\usepackage{silence} % silence warnings
\usepackage[caption=false]{subfig}
\usepackage{svg}

\hypersetup{
    colorlinks = true,
    linkcolor = black, % blue,
    urlcolor = black, % blue,
    citecolor = black % Makes citations look like regular text
}

\usepackage{tikz}
\usetikzlibrary{positioning, arrows.meta, calc}
\newtheorem{property}{Property}

\newcommand{\ii}{\text{i}}

\newcommand{\Imag}{\text{Im}}

\newcommand{\chiPattern}[2]{\chi_{#1}^{(#2)}}
 
\newcommand{\chiConjugatePattern}[2]{(\chi_{#1}^{(#2)})^{*}}
\newcommand{\cTilde}{\Tilde{c}}

\begin{document}
\title{Learning rules for complex-valued patterns in networks of oscillators}
\author{Federico Sbravati\,\orcidlink{0009-0008-2079-7658}
}
% \email{f.sbravati@tue.nl}
%\altaffiliation{NanoComputing Research Lab, Integrated Circuits Group, Electrical Engineering Department}

\author{Aida Todri-Sanial\,\orcidlink{0000-0001-8573-2910}
}
% \thanks{Corresponding author:}
\email{a.todri.sanial@tue.nl}
%\altaffiliation{NanoComputing Research Lab, Integrated Circuits Group, Electrical Engineering Department}
\affiliation{NanoComputing Research Lab, Integrated Circuits Group,\\Department of Electrical Engineering, Eindhoven University of Technology, De Groene Loper 19, 5612 AP Eindhoven, The Netherlands}

% \date{\today}% It is always \today, but any date may be explicitly specified
\date{August 22, 2026}% It is always \today, but any date may be explicitly specified

%%%%%%%%%%% Section input
 % Use showkeys class option if keyword display desired

\begin{abstract}
    In this article, we extend learning rules from real binary to complex-valued spins.
    This formulation allows for a robust and natural representation of grayscale patterns, where spins behave as multi-state neurons and can be stored in a complex-valued weight matrix.
    We describe a rule that performs better than standard methods, such as Hebbian learning, to encode information in a suitable form for pattern retrieval with networks of oscillators.
    Since in neural networks it is of interest to have local and incremental learning rules, we prove the extension of a result by Diederich and Opper with our complex-valued spin formulation.
    We then test the behavior of the associative memory for the system of oscillators under different circumstances for both real-valued, as well as complex-valued correlated and random patterns.
\end{abstract}

\maketitle
\section{Introduction}\label{sec:02_introduction}
The Kuramoto model \cite{kuramoto_formation_1975, mori_dissipative_1998} has been extensively studied in many areas of physics to investigate the dynamics of synchronization phenomena \cite{daido_order_function_1992, DAIDO1993394, daido_generic_scaling_1994, DAIDO199624, daido_algebraic_2000} in biology \cite{Ermentrout1991adaptive}, photonics \cite{Takemura2021Emulating}, mechanics \cite{ebrahimzadeh_minimal_2020, ebrahimzadeh_mixed-mode_2022}, power grid modeling \cite{olmi_hysteretic_2014}, populations of oscillators \cite{floriach_chimeras_2025}, and Josephson Junctions \cite{wiesenfeld_synchronization_1996}, among many other disciplines and applications.
Moreover, phase-based networks of coupled oscillators have recently been used to describe and develop computational frameworks for associative memory and combinatorial optimization \cite{todri-sanial_how_2022, haverkort_solving_2026, delacour_lagrange_2025}.

The Hopfield model \cite{Hopfield_1982} is another influential model in physics and neural-network theory, particularly for the study of associative memory in binary spin networks.
Both of these disciplines can be combined in the field of oscillator networks \cite{hoppensteadt_pattern_2000, todri-sanial_computing_2024}, where it is possible to encode information in the relative phase relationships of the coupled oscillators.
Patterns can be binary in the case of a real-valued coupling matrix $J_{ij}$ (similar to the classical Hopfield network) or multi-valued, i.e. \textit{grayscale}, just like in the continuous version of the Hopfield networks \cite{1984_Hopfield_neurons_graded_response}, in the case of complex-valued couplings $J_{ij} = |J_{ij}| e^{\ii \arg(J_{ij})}$.
In the first case information is only encoded in the amplitude of the couplings, whereas in the complex case it is encoded in both amplitude and phase of the weight matrix.
As a result, the evolution of the system to a target pattern depends both on the relative strength and phase between pairs of Kuramoto oscillators.

In recent years there have been efforts to improve capacity by dense networks \cite{krotov_dense_2016}, which have been adapted to the Kuramoto model \cite{2025_berloff_higher_order_kuramoto_oscillator_network} using Hebbian learning with higher order interactions. 
Here we report on an extension of learning rules from binary to grayscale patterns, to efficiently retrieve them when using Kuramoto networks under the condition of finite memory load for both structured and random sets of patterns.

The Pseudoinverse, or Moore-Penrose inverse \cite{penrose_generalized_1955}, is a generalized notion of a matrix inverse and can be defined for of neural networks to encode the weights for patterns \cite{kanter_associative_1987}.
The properties of this matrix allow it to be extended to the field of complex numbers $\mathbb{C}$ \cite{penrose_generalized_1955} and therefore applied to complex-valued patterns.
A local approximation that converges to the Pseudoinverse was introduced by Diederich and Opper in 1987 \cite{diederich_learning_1987}, which we adapt in this work to learn complex patterns.
Thus, the spins $\{\sigma_{i}^{(\mu)}\}_{\mu = 1}^{P} $ encoding $P$ patterns are not binary, but can take up multiple values.
The concept of Pseudoinverse has found applications that are important for pattern retrieval \cite{kanter_associative_1987}, MRI \cite{yeung_algebraic_2025}, photonics \cite{2026_Silva_photonic_learning_machines}, which is why it is a worthwhile mathematical construction to investigate in conjunction with the dynamical properties of the Kuramoto model.

% Sections of the paper
This work is divided up into the following sections.
The introduction in Section \ref{sec:03_background} gives the relevant background information on neural networks and Kuramoto oscillators.
In Section \ref{sec:04_methods} we introduce the physical quantities we use to quantify the retrieval of patterns as a function of initial noise applied to the dataset.
Moreover, we extend the learning rules from how they are used in binary Hopfield networks \cite{Hopfield_1982} to continuous complex data encoding for the Kuramoto model.
In Section \ref{sec:05_simulations} we analyze the simulations that have been carried out to validate the encoding and retrieval by means of the dynamics of Kuramoto oscillator networks.
Finally, in Section \ref{sec:06_conclusions} we discuss the results of the paper and present our conclusions.
\section{Background}\label{sec:03_background}
% \textreview{Basically cite our relevant stuff.}
% ////
In neural network theory the most well-known approach to encode patterns (or memories) in a matrix is provided by the \textit{Hebbian learning rule} \cite{1949_Hebb_organization_behavior}, which is defined by the outer product 
\begin{equation}
    J_{ij}^{(\text{Hebb})} := \dfrac{1}{P}\sum_{\mu = 1}^{P} \chiPattern{i}{\mu}\chiPattern{j}{\mu} \,\,\,
\end{equation}
of $P$ (real) patterns $\boldsymbol{\chi} \in \mathbb{R}^{N}$.
Then, through some choice of dynamical process, it is possible to retrieve the memories them from a partially corrupted state.
Out of the possible choices the most well-known kind of network dynamics are the ones laid out by Hopfield in 1982 \cite{Hopfield_1982}, where he quantified and described the retrieval properties of a binary spin network to retrieve memories using the Hebbian rule.
Oscillator (neural) networks (ONN) have also been an important subject of study in the context of dynamical systems theory and applications.
The reason for this is that starting from the concept of \textit{synchronization}, where the sole evolution of coupled oscillators takes place, it is possible to describe how a collection of these is able to show emergent phenomena \cite{olmi_hysteretic_2014, ebrahimzadeh_mixed-mode_2022, floriach_chimeras_2025}. 
% (\textreview{rephrase, but concept is there. Put citations like Olmi, Ebrahimzadeh, etc.}).
Moreover, it is possible to exploit the synchronizing behavior to encode information into the couplings for associative memory problems \cite{todri-sanial_computing_2024}, as well as optimization problems \cite{haverkort_solving_2026}.

The most popular model to describe the phase dynamics of weakly coupled synchronizing oscillators, is the one first introduced by Kuramoto in 1975 \cite{kuramoto_formation_1975}
\begin{equation}\label{eq:kuramoto_model_real}
    \dot{\varphi}_{i} = \omega^{(0)}_{i} - \sum_{j = 1}^{N} J_{ij} \sin(\varphi_{i}^{(\mu)} - \varphi_{j}^{(\mu)}) \,\,\,,
\end{equation}
where $\omega_{i}^{(0)}$ are the different natural frequencies of the oscillators.
This ordinary differential equation (ODE) can also be interpreted as the as gradient descent of the Hamiltonian $\mathcal{H}_{\text{XY}}$ of the XY model \cite{kosterlitz_ordering_1973, kosterlitz_critical_1974} when $\omega_{i}^{(0)} = c$ for all oscillators $i$
\begin{equation}
    \Dot{\varphi}_{i} = \left(-\dfrac{\partial \mathcal{H}_{\text{XY}}}{\partial \varphi_{i}} \right)_{\omega_{i}^{(0)} = c} \,\,\, ,
\end{equation}
for a constant $c$.
Since the frequencies under this assumption become a constant of integration, under a change of frame of reference they can then be neglected in the dynamical evolution of the system.

\section{Methods}\label{sec:04_methods}
We first introduce the mapping methodology to go from grayscale patterns to the complex domain.
Then, we present the definitions of the physical quantities needed to quantify our results.
These are then linked them with standard metrics in Kuramoto theory, such as the order parameter that measures global \textit{synchronization} \cite{mori_dissipative_1998, daido_algebraic_2000, strogatz_kuramoto_2000}.
Moreover, we merge this notion with the notion of \textit{overlap} \cite{1980_Parisi_order_parameter_spin_glasses} from spin glass theory that has found use in neural networks \cite{kanter_associative_1987, krauth_learning_1987, agliari_networks_2026}, which we extend to complex spins to quantify the similarity of complex patterns.

\subsection{Mapping of patterns to phases}\label{ssec:04a_mapping}
To ensure that pattern retrieval correctly works under the dynamics of Equation \eqref{eq:kuramoto_model_real}, we need to define an appropriate mapping.
This will be the first step in defining the associative memory properties of the Kuramoto model.
At the core of this, is the notion that for a physical system to compute through natural dynamics the proper mapping has to be found \cite{wolpert_what_2026, haverkort_solving_2026}.
We consider first a black-and-white image to be represented by $0$ (black) and $1$ (white).
In the binary case a pixel $x_{i}^{(\mu)} \in \{0,1\}$, whereas on a scale that goes from black to white (both with discrete or continuous values) the range of the pixel will also take the values in-between $x_{i}^{(\mu)} \in [0,1]$.
For a grayscale image levels between the two extrema are mapped accordingly, through a linear function.
The mapping from pixels $x_{i}^{(\mu)}$ to phases is defined by the following formula
\begin{equation}\label{def:pixel_phase_mapping}
    \alpha_{i}^{(\mu)} := \pi x_{i}^{(\mu)} \,\,\, ,
\end{equation}
where $x_{i}^{(\mu)} \in [0, 1]$ is the domain of pixels.

\begin{figure}[htbp]
    \centering
    \begin{tikzpicture}
    % Radius
    \def\r{1.5}
    \def\rext{1.62} % slightly outside the circle

    % Draw unit circle
    % \draw[thick] (0,0) circle(\r);
    \draw[thick] (-\r,0) arc[start angle=180,end angle=0,radius=\r];

    % Draw axes
    \draw[->] (-2.0, 0.0) -- (2,0) node[right] {Re};
    \draw[->] (0.0, 0.0) -- (0,2) node[above] {Im};

    % Points on the circle
    \filldraw[black] ({\r*cos(0)},{\r*sin(0)}) circle(2pt) node[below right] {$(1; 0)$};
    \filldraw[black] ({\r*cos(35)},{\r*sin(35)}) circle(2pt) node[above right] {$(2;  \alpha_{2})$};
    \filldraw[black] ({\r*cos(80)},{\r*sin(80)}) circle(2pt) node[above right] {$(3;  \alpha_{3})$};
    \filldraw[black] ({\r*cos(140)},{\r*sin(140)}) circle(2pt) node[above left] {$(n-1; \alpha_{n-1})$};
    \filldraw[black] ({\r*cos(180)},{\r*sin(180)}) circle(2pt) node[below left] {$(n; \pi)$};
    % \filldraw[black] ({\r*cos(235)},{\r*sin(235)}) circle(2pt) node[below left] {$6$};
    % \filldraw[black] ({\r*cos(275)},{\r*sin(275)}) circle(2pt) node[below right] {$7$};

    % Continuation dots slightly outside the circle, along the same arc
    \fill ({\rext*cos(105)},{\rext*sin(105)}) circle(0.8pt);
    \fill ({\rext*cos(110)},{\rext*sin(110)}) circle(0.8pt);
    \fill ({\rext*cos(115)},{\rext*sin(115)}) circle(0.8pt);

    % Final point labeled n
    % \filldraw[black] ({\r*cos(310)},{\r*sin(310)}) circle(2pt) node[below right] {$n$};

    % Arrow from center to the n-th point
    \draw[->, thick] (0,0) -- ({\r*cos(140)},{\r*sin(140)});
    \node at ({0.82*cos(120)},{0.95*sin(120)}){$ r = 1 $}; % node[below, left]{$ r_{ij}$};
    
    % Angle arc
    \draw[thick] (0.45,0) arc[start angle=0,end angle = 140,radius = 0.45];
    \node at ({0.7*cos(50)},{0.7*sin(50)}) {$\alpha_{n - 1}$};
\end{tikzpicture}
    \caption{
        This is a conceptual representation of how the phase values are mapped onto the circle with radius one, similarly to how the Kuramoto model has been mapped to encode multi-state values on the unit circle \cite{haverkort_solving_2026}.
        Overall the radius and angle of the circle represent the $n$ grayscale levels one can encode in the gradient.
    }
    \label{fig:unit-semicircle-pixels-to-phases}
\end{figure}
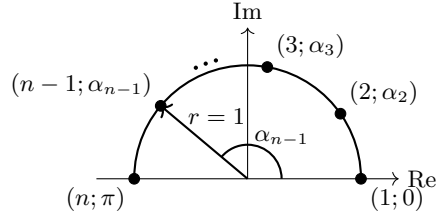

Then, wanting to represent pixels as phases we rescale $x_{i}^{(\mu)}$, where $x_{i}^{(\mu)} \mapsto \alpha_{i}^{(\mu)}$ and is shown in Figure \ref{fig:unit-semicircle-pixels-to-phases}.
Thus, we defined a mapping to compute with our dynamical system \cite{wolpert_what_2026, haverkort_solving_2026} through a function $f$
\begin{equation}\label{eq:mapping_phases_to_complex_patterns}
    f(\{ x_{i}^{(\mu)} \}) =  e^{\ii \pi x_{i}^{(\mu)}}  \,\,\, .
\end{equation}
Formally the mapping is a function that takes pixels from their domain to be represented as elements of the unitary group $U(1)$
\begin{equation}
    f: \Omega_{x} = [0, 1] \mapsto U(1) = \chiPattern{i}{\mu} \,\,\, .
\end{equation}
Thus, we define the \textit{complex patterns} using the following notation
\begin{equation}\label{def:complex_spin_pattern}
    \chiPattern{i}{\mu} := e^{\ii \alpha_{i}^{(\mu)}} \,\,\, ,
\end{equation}
where $\chiPattern{i}{\mu}$ is the pattern and $\alpha_{i}^{(\mu)}$ are the values represented as phases for the $i$-th oscillator and $\mu$-th pattern.

Finally, we also need a \textit{readout} mapping for the steady-state phases defined through $g$
\begin{equation}\label{eq:mapping_inverse_complex_patterns_to_phases}
    g: [0, \pi] \mapsto [-1, 1] \,\,\, ,
\end{equation}
where the function we use is
\begin{equation}
    g(\varphi) := \cos \left( \varphi_{i}^{(\mu)}(\infty) \right) \,\,\, ,
\end{equation}
with $\varphi_{i}^{(\mu)}(\infty)$ being the phase of a pattern at steady-state $t^{*} = \infty$.
It is important to remark that all patterns are stable points of the learned weight matrix and as complex patterns possess a global rotational symmetry by an arbitrary rotation $\vartheta$
\begin{equation}
   \chi_{i}^{(\mu)} \mapsto \chi_{i}^{(\mu)} e^{\ii \vartheta} \,\,\, , \,\,\, \forall i , \forall \mu \,\,\, .
\end{equation}

% Potts mapping
An analogous mapping in terms of extension of the Ising model can be applied to the clock model \cite{Potts_1952} with a similar approach as Equation \eqref{def:complex_spin_pattern}.
Other mappings can also be found to match the $q$-state Potts model, where $q = \{1, \dots, n_{L}\}$ for $n_{L}$ states.
%
%
% ANALYTICAL %
\subsection{Analytical methods}\label{ssec:04b_analytical}
In this section we define the relevant physical quantities, then introduce the notion of the Pseudoinverse learning rule with complex patterns, and finally prove that rule II from Diederich and Opper \cite{diederich_learning_1987} also holds for complex patterns and coincides with the coupling matrix constructed using the Pseudoinverse construction \cite{kanter_associative_1987}.

\subsubsection{Physical quantities}\label{sssec:04b_1_physical_quantities}
The \textit{order parameter} we use to quantify the similarity of patterns in a dataset for oscillator networks is the \textit{complex overlap} $\Psi_{\mu\nu}$, which we define \footnote{
    We borrow the symbols from the theory of superconductivity by Landau and Ginzburg, due to the resemblance of oscillator order parameters having an amplitude and a global average phase to the order parameter of the Landau-Ginzburg theory of superconductivity.
    This choice is grounded in the use of the Kuramoto model has also been used to describe Josephson junction arrays \cite{wiesenfeld_synchronization_1996}.
} as 
\begin{equation}\label{def:complex_overlap_order_parameter}
    \Psi_{\mu\nu} = |\psi_{\mu\nu}| e^{\ii \Phi_{\mu\nu}} := \dfrac{1}{N} \sum_{i = 1}^{N} e^{\ii ( \varphi_{i}^{(\mu)} - \varphi_{i}^{(\nu)} )} \,\,\, ,
\end{equation}
where $N$ is the number of oscillators, $\Phi_{\mu\nu}$ the global phase and $|\psi_{\mu \nu}|$ the global absolute value between pairs of patterns $\mu$ and $\nu$.

We substitute in the subsequent text $|\psi_{\mu\nu}|$ with the conventional symbol for synchronization in the Kuramoto model, namely $\rho_{\mu\nu}$.
Similarly, the definition of the \textit{complex overlap} as an order parameter is directly inspired by spin-glass theory \cite{sherrington_solvable_1975, 1980_Parisi_order_parameter_spin_glasses}.
We define the absolute value of the complex order parameter as the \textit{synchronization (overlap)}, which in our case is the matrix formulation of the commonly used order parameter to quantify the global synchronization of the oscillator network \cite{daido_order_function_1992}.
It is taken as the average over the experiments at a fixed value of initial noise $\sigma$, which implies that $\Psi_{\mu\nu}(\sigma)$ changes as a function of the standard deviation in the initial condition.
Moreover, it is important to note that $\Psi_{\mu\nu}(\sigma)$ is normalized over the number of patterns and thus is defined as:
\begin{equation}\label{def:synchronization_overlap_order_parameter}
    \rho_{\mu\nu}(\sigma) := |\Psi_{\mu\nu}(\sigma)| \,\,\, ,
\end{equation}
We define the \textit{normalized trace} of the synchronization overlap as the quantity defined as:
\begin{equation}\label{def:normalized_synchronization_overlap_trace}
    R := \text{tr}_{\text{P}} \lbrace \rho_{\mu \nu} \rbrace = \dfrac{1}{P}\sum_{\mu = 1}^{P} \rho_{\mu\mu} \,\,\, ,
\end{equation}
where $P$ is the number of memorized patterns and $R$ takes values in $[0, 1]$, which can be computed for each level of initial pattern noise $\sigma$.
This is physically equivalent to the \textit{synchronization} in standard Kuramoto theory \cite{mori_dissipative_1998,strogatz_kuramoto_2000, ott_low_2008}.
% Similarly to the commonly used synchronization in the Kuramoto model the range of $R$ is $[0, 1]$.
For $R = 0$ there is no synchronization or similarity of the patterns, whereas for $R = 1$ the patterns coincide.
In practical settings for similar patterns one only has full synchronization where $R \rightarrow 1$.

\subsubsection{Complex Pseudoinverse}\label{sssec:04b_2_complex_pseudoinverse}
\begin{figure}[htbp]
    \centering
    % OLD - FULL CIRCLE
\begin{tikzpicture}
    % Radius
    \def\r{1.5}
    \def\rext{1.62} % slightly outside the circle

    % Draw unit circle
    \draw[thick] (0,0) circle(\r);

    % Draw axes
    \draw[->] (-2,0) -- (2,0) node[right] {Re};
    \draw[->] (0,-2) -- (0,2) node[above] {Im};

    % Points on the circle
    \filldraw[black] ({\r*cos(0)},{\r*sin(0)}) circle(2pt) node[below right] {$(1; 0)$};
    \filldraw[black] ({\r*cos(35)},{\r*sin(35)}) circle(2pt) node[above right] {$2$};
    \filldraw[black] ({\r*cos(80)},{\r*sin(80)}) circle(2pt) node[above right] {$3$};
    \filldraw[black] ({\r*cos(140)},{\r*sin(140)}) circle(2pt) node[above left] {$4$};

    % Continuation dots slightly outside the circle, along the same arc
    \fill ({\rext*cos(150)},{\rext*sin(150)}) circle(0.8pt);
    \fill ({\rext*cos(155)},{\rext*sin(155)}) circle(0.8pt);
    \fill ({\rext*cos(160)},{\rext*sin(160)}) circle(0.8pt);
    
    \filldraw[black] ({\r*cos(180)},{\r*sin(180)}) circle(2pt) node[below left] {$(k; \pi)$};
    \filldraw[black] ({\r*cos(215)},{\r*sin(215)}) circle(2pt) node[below left] {$k + 1$};
    \filldraw[black] ({\r*cos(260)},{\r*sin(260)}) circle(2pt) node[below left] {$k + 2$};

    % Continuation dots slightly outside the circle, along the same arc
    \fill ({\rext*cos(290)},{\rext*sin(290)}) circle(0.8pt);
    \fill ({\rext*cos(295)},{\rext*sin(295)}) circle(0.8pt);
    \fill ({\rext*cos(300)},{\rext*sin(300)}) circle(0.8pt);

    % Final point labeled n
    \filldraw[black] ({\r*cos(310)},{\r*sin(310)}) circle(2pt) node[below right] {$n$};

    % Arrow from center to the n-th point
    \draw[->, thick] (0,0) -- ({\r*cos(310)},{\r*sin(310)}) node[midway, right]{$ |J_{ij}|$};
    
    % Angle arc
    \draw[thick] (0.45,0) arc[start angle=0,end angle= -50,radius=0.45];
    \node at ({0.75*cos(25)},{0.75*sin(25)}) {$\arg(J_{ij})$};
\end{tikzpicture}
    \caption{
        This is a conceptual representation of how the weight matrix values $J_{ij}$ are distributed onto the circle.
        Overall the radius $|J_{ij}|$ and angle $\arg(J_{ij})$ of the circle represent the $n$ encoded grayscale levels one can encode in the gradient.
        The dynamics in the Kuramoto model are equivalent to a term controlled in amplitude and phase $|J_{ij}| \sin\left[\varphi_{i} - \varphi_{j} + \arg(J_{ij}) \right]$, where $|J_{ij}|$ is the strength of the coupling and $\arg(J_{ij})$ is the relative angle between two oscillators.
    }
    \label{fig:unit-circle-weight-matrix}
\end{figure}
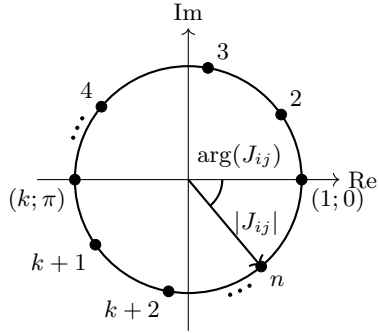

First, we define the coupling matrix defined by the Pseudoinverse construction \cite{kanter_associative_1987}, in the context of complex patterns as defined in Equation \eqref{def:complex_spin_pattern}.
A conceptual representation of how the patterns are stored both in the absolute value as well as in the angle part of the weight matrix is shown in Figure \ref{fig:unit-circle-weight-matrix}.
The grayscale image is mapped such that both the strength $|J_{ij}|\in [0, 1]$, as well as the relative angle between two oscillators control the evolution of the phases.
Given the matrix of complex patterns $X_{\mu i} \in \mathbb{C}^{P \times N}$, where $P$ is the number of patterns and $N$ is the total number of oscillators, which for a two-dimensional patterns is the product of all pixels, namely $N = N_{x} N_{y}$.
As a result, the weight matrix that stores the patterns through a \textit{complex Pseudoinverse} is defined in analogy to the real spin formulation \cite{kanter_associative_1987} as
\begin{equation}\label{def:pseudoinverse_complex}
     J_{ij}^{(\text{Pseudo})} :=
        \sum_{\mu,\nu = 1}^{P} \chiConjugatePattern{i}{\mu} (\Psi^{-1})_{\mu\nu} \chiPattern{j}{\nu} \,\,\, .
\end{equation}
The matrix $\boldsymbol{\Psi}$ is the \textit{complex overlap} order parameter defined in equation \eqref{def:complex_overlap_order_parameter}.
Alternatively the coupling matrix can be defined equivalently as
\begin{equation}\label{def:pseudoinverse_complex_alt}
     J_{ij}^{(\text{Pseudo})} := X^{+} X \,\,\, ,
\end{equation}
where $X^{+} \in \mathbb{C}^{N \times P}$ is the Moore-Penrose generalized inverse \cite{penrose_generalized_1955}, also known as \textit{Pseudoinverse} of the pattern matrix, defined as $X^{+} = X^{\dagger} (X X^{\dagger})^{-1}$, where the complex overlap from Equation \eqref{def:complex_overlap_order_parameter} in this notation is $\mathbf{\Psi} = X X^{\dagger}$ and $X^{\dagger}$ denotes the Hermitian conjugate of the pattern matrix.

\subsubsection{Local Pseudoinverse}\label{sssec:04b_3_local_pseudoinverse}
We want to show that the construction by Diederich and Opper \cite{diederich_learning_1987} can be extended to patterns in the complex form as we defined them in Equation \eqref{def:complex_spin_pattern}.
For a correspondence between the symbols used in this paper and the one by Diederich and Opper \cite{diederich_learning_1987} a table can be found in Appendix \ref{sec:appendix_a_translation}.
We assume the patterns $\chi_{i}^{(\mu)}$ to be linearly independent $\forall \mu$.
We define the Hermitian conjugate as the vector
\begin{equation}\label{def:pattern_hermitian}(\boldsymbol{\chi}^{(\mu)})^{\dagger} := (e^{-\ii \alpha_{1}^{(\mu)}}, \dots, e^{-\ii \alpha_{N}^{(\mu)}}) \,\,\, .
\end{equation}
We start by defining the local field for complex patterns as an extension to binary spins \cite{kanter_associative_1987, diederich_learning_1987}:
\begin{equation}\label{def:local_field_complex}
    h_{i}^{(\mu)} := \sum_{j = 1}^{N} J_{ij} \chiConjugatePattern{j}{\mu} \,\,\, , \,\,\, \forall \mu \,\,\, ,
\end{equation}
which we use to define the \textit{embedding condition}, which when satisfied it implies that the Pattern $\chiPattern{i}{\mu}$ is a stable fixed point and has thus been memorized by the system.
The \textit{embedding condition} in the complex case to store the patterns as stable minima of the network  \cite{kanter_associative_1987, diederich_learning_1987} is
\begin{equation}\label{eq:stability_condition}
     \chiPattern{i}{\mu} h_{i}^{(\mu)} = \sum_{j = 1}^{N} \chiPattern{i}{\mu} J_{ij} \chiConjugatePattern{j}{\mu} > 0 \,\,\, .
\end{equation}
A correctly embedded pattern $\chiPattern{i}{\mu}$ therefore implies that 
\begin{equation}\label{eq:embedding_stability_condition}
     \chiPattern{i}{\mu} h_{i}^{(\mu)} = 
    \chiPattern{i}{\mu}\sum_{j = 1}^{N} J_{ij} \chiConjugatePattern{j}{\mu} = 1 \,\,\, 
\end{equation}
holds and thus the embedding condition of Equation \eqref{eq:embedding_stability_condition} is satisfied.
In analogy with the derivation from \cite{diederich_learning_1987} the update per pattern $\delta J_{ij}^{(\mu)}$ is defined as
\begin{equation}
    \delta J_{ij}^{(\mu)} := \dfrac{1}{N} \left[ 1 - \chiPattern{i}{\mu} h_{i}^{(\mu)} \right] \xi_{ij}^{(\mu)}\,\,\, .
\end{equation}
\\
Since we have that for each neuron $i$ there is a coefficient that can be fixed at each learning step $l$, for every pattern $\mu$, we expand then the matrix along the \textit{effective pattern direction} $\xi_{ij}^{(\mu)}$, which is defined as follows
\begin{equation}\label{def:xi_effective_pattern_direction}
    \xi_{ij}^{(\mu)} := \chiConjugatePattern{i}{\mu} \chiPattern{j}{\mu} \,\,\, .
\end{equation}
\\
The definition follows the convention used by Diederich and Opper \cite{diederich_learning_1987} we use the symbol $\xi_{ij}^{(\mu)}$, due to the need of extending the treatment to complex values.
Expanding the previous term, we then find 
\begin{equation}\label{eq:update_l_to_l+1_step}
    J_{ij} \rightarrow J_{ij} + \dfrac{1}{N} \Bigg[ 1 -  \sum_{k = 1}^{N} J_{ik} \left(\xi_{ik}^{(\mu)}\right)^{*} \Bigg]\xi_{ij}^{(\mu)} \,\,\, .
\end{equation}

This is therefore coherent with the definition of inner product for complex patterns and justifies the chosen nomenclature.
Substituting the definition from Equation \eqref{def:xi_effective_pattern_direction} into the weight matrix expansion we find
\begin{equation} 
    J_{ij} = \dfrac{1}{N}\sum_{\nu = 1}^{P} c_{i}^{(\nu)} \xi_{ij}^{(\nu)} =
    \dfrac{1}{N}\sum_{\nu = 1}^{P} c_{i}^{(\nu)}\chiConjugatePattern{i}{\nu} \chiPattern{j}{\nu} \,\,\,.
\end{equation}
We use the complex analog $c_{i}^{(\mu)}(l)$ of the embedding strengths $x^{(\mu)}(l)$ used in the original paper \cite{diederich_learning_1987}, where the coefficients were assumed to be real valued for rule II by the authors \footnote{Moreover, if the diagonal $J_{ii}$ were taken to be zero, the sufficient conditions for convergence would have to be more stringent. As a result all the calculations would have to be solved and hold for every oscillator $i$, namely every coefficient $c_{i}(\infty)$ used for reconstruction would require to solve the system of equations and thus have invertible reconstruction matrix $\Gamma_{i}$ that simultaneously depends on all other coupled oscillators excluding the $i$-th.}.
Moreover, for the proof we keep the index $i$ for the expansion of the matrix explicit, but assume it fixed as in \cite{diederich_learning_1987} and then to hold for all incoming connection of each neuron $i$.

Then, using the right-hand side of equation \eqref{eq:update_l_to_l+1_step} for $c_{i}^{(\mu)}(l + 1)$ and further substituting the expansion of $J_{ij}$ for the $i$-th neuron, we obtain for the $(l + 1)$ learning step 
\begin{widetext}
    \begin{equation}\label{eq:finite_difference_pattern_coefficients_i}
        c_{i}^{(\mu)}(l + 1) - c_{i}^{(\mu)}(l) = 
        1 - \chiPattern{i}{\mu}\sum_{k = 1}^{N} J_{ik} \chiConjugatePattern{k}{\mu} = 
        1 - \dfrac{1}{N} \Bigg\{ \sum_{\nu = 1}^{P} \sum_{k = 1}^{N} c_{i}^{(\nu)} \chiConjugatePattern{i}{\nu} \chiPattern{k}{\nu} \chiPattern{i}{\mu} \chiConjugatePattern{k}{\mu}  \Bigg\} \,\,\, .
    \end{equation}
\end{widetext}
On the right-hand side of Equation \eqref{eq:finite_difference_pattern_coefficients_i} we have that all coefficients for 
\begin{itemize}
    \item $\nu < \mu$ have been updated s.t. $c_{i}^{(\mu)} = c_{i}^{(\mu)}(l + 1) $;
    \item $\nu \geq \mu$ have \textit{not} been updated s.t. $c_{i}^{(\mu)} = c_{i}^{(\mu)}(l) $.
\end{itemize}
Therefore, we get
\begin{subequations}\label{eq:finite_difference_coefficients_learning_steps}
\begin{align}
    &c_{i}^{(\mu)}(l + 1) - c_{i}^{(\mu)}(l) = 1 - \sum_{\nu < \mu} \Gamma_{i}^{\mu \nu} c_{i}^{(\nu)} - \sum_{\nu \geq \mu} \Gamma_{i}^{\mu \nu} c_{i}^{(\nu)} \label{eq:coefficients_step_a}
    \\
    &\implies c_{i}^{(\mu)}(l + 1) = 1 - \sum_{\nu < \mu} \Gamma_{i}^{\mu \nu} c_{i}^{(\nu)} - \sum_{\nu > \mu} \Gamma_{i}^{\mu \nu} c_{i}^{(\nu)}
    \,\,\, ,\label{eq:coefficients_step_b}
\end{align}
\end{subequations}
where the $c_{i}^{(\mu)}(l)$ on the left-hand and right-hand side canceled out between the two steps (we separated out the second sum between $\nu = \mu$ and $\nu > \mu$ and simplified $c_{i}^{(\mu = \nu)}(l)$ with the right-hand side).
The symmetric matrix $\Gamma_{i}^{\mu\nu}$ is defined as 
\begin{equation}\label{def:matrix_gamma_definition}
    \Gamma_{i}^{\mu \nu} := \dfrac{1}{N}\sum_{k = 1}^{N} \left(\xi_{ik}^{(\mu)}\right)^{*}\xi_{ik}^{(\nu)} \,\,\,,
\end{equation}
which when expanded yields
\begin{equation}\label{eq:matrix_gamma_expanded}
    \Gamma_{i}^{\mu \nu} = \dfrac{1}{N}\sum_{k = 1}^{N} \chiPattern{i}{\mu} \chiConjugatePattern{k}{\mu} \chiConjugatePattern{i}{\nu}\chiPattern{k}{\nu} \,\,\, .
\end{equation}
Assume now that the limit for $l \rightarrow \infty$ exists for every pattern $\nu \in \{1, \dots, P \}$.
Then the following linear system of equation is satisfied (follows from applying the limit from the previous assumption to equation \eqref{eq:finite_difference_coefficients_learning_steps})
\begin{equation}\label{eq:gauss_seidel_method_convergence}
    \sum_{\nu = 1}^{P} \Gamma_{i}^{\mu\nu} c_{i}^{(\nu)}(\infty) = 1 \,\,\, ,
\end{equation}
where Equation \eqref{eq:gauss_seidel_method_convergence} is the Gauss-Seidel method \cite{quarteroni_numerical_2017} used to solve the linear system from Equation \eqref{eq:finite_difference_coefficients_learning_steps}.
It converges if $\Gamma_{i}^{\mu\nu}$ is positive definite for all $i$ \cite{diederich_learning_1987}, 
namely if
\begin{equation}\label{eq:quadratic_form}
    \sum_{\mu, \nu} (\Tilde{c}_{i}^{(\mu)})^{*} \Gamma_{i}^{\mu\nu} \Tilde{c}_{i}^{(\nu)} = \dfrac{1}{N} \sum_{j = 1}^{N} \Bigg|\sum_{\mu = 1}^{P} \Tilde{c}_{i}^{(\mu)} \chiConjugatePattern{i}{\mu} \chiPattern{j}{\mu} \Bigg|^{2} > 0 \,\,\, ,
\end{equation}
is \textit{strictly positive} for each pattern/configuration, some coefficients $\Tilde{c}_{i}^{(\mu)}$, for every $\mu \in \{1, \dots, P\}$ that does not vanish identically.
This condition is satisfied, because we initially assumed the patterns $\chiPattern{i}{\mu}$ to be linearly independent.

Now we show that we can reconstruct the $J_{ij}$ matrix with the row-wise expansion, once the system is at convergence by solving Equation \eqref{eq:gauss_seidel_method_convergence}.
To do this, we take the matrix $\Gamma_{i}^{\mu\nu}$ to be defined by the effective pattern directions, at a fixed oscillator $i$ and show that \eqref{eq:quadratic_form} holds.
First, we have that for a pattern $\chi_{i}^{(\mu)}$ the expansion in terms of the coefficients $c_{i}^{(\mu)}$ to reconstruct the patterns, yields
\begin{equation}\label{eq:expansion_gamma_pattern}
    \sum_{\nu} \Gamma_{i}^{\mu\nu} c_{i}^{(\nu)} = 1 \,\,\, .
\end{equation}
Then we recognize that Equation \eqref{eq:matrix_gamma_expanded} can be rearranged and rewritten to incorporate the complex overlap $\Psi$ as
\begin{equation}
    \Gamma_{i}^{\mu\nu} = \chiPattern{i}{\mu} \chiConjugatePattern{i}{\nu} \Psi_{\nu\mu} \,\,\, .
\end{equation}

At this point we introduce the diagonal unitary matrix $U_{i}^{\mu\nu}$
\begin{equation}\label{eq:unitary_matrix_pattern_diagonal}
    U_{i}^{\mu\nu} := \chiPattern{i}{\mu} \delta_{\mu\nu} \,\,\, ,
\end{equation}
where $\delta_{\mu\nu}$ is the Kronecker delta and $\Psi_{\nu\mu}$ is the conjugate of the complex overlap.
We keep the oscillator index explicit to then show the validity of Equation \eqref{eq:gauss_seidel_method_convergence}. 
Now we exploit the expansion of $\Gamma$ from equation \eqref{eq:expansion_gamma_pattern} to invert the equation and solve explicitly for the coefficients
\begin{equation}\label{eq:c_strength_inverse_gamma}
    c_{i}^{(\mu)} = \sum_{\nu = 1}^{P} (\Gamma_{i}^{-1})_{\mu\nu} \,\,\, .
\end{equation}
Since for a fixed pattern we have that the unitary matrix defined in Equation \eqref{eq:unitary_matrix_pattern_diagonal} applied to a pattern yields the identity
\begin{equation}
    (U_{i}^{\dagger})^{\mu\nu}\chiPattern{i}{\mu} = \chiConjugatePattern{i}{\mu} \chiPattern{i}{\mu} = 1 \,\,\,.
\end{equation}
From the previous steps it is important to keep in mind that the following relationship holds
\begin{equation}
    \Gamma_{i}^{\mu\nu} = \sum_{\alpha, \beta} U_{i}^{\alpha\mu}\Psi_{\beta\alpha} (U_{i}^{\dagger} )^{\beta\nu}\,\,\, ,
\end{equation}
which since $U$ is unitary implies
\begin{equation}
    (\Gamma_{i}^{-1})_{\mu\nu} = U_{i}(\Psi^{-1})_{\nu\mu} U_{i}^{\dagger} \,\,\, .
\end{equation}
This remark in turn implies that for each coefficient $c_{i}^{(\mu)}$ the following holds
\begin{subequations}    \begin{align}\label{eq:coefficient_reconstructed_solution}
        c_{i}^{(\mu)} &= \sum_{\nu} (\Gamma_{i}^{-1})_{\mu\nu} =
        \sum_{\nu,\eta}
        (U_{i}^{\dagger})_{\mu\nu}
        (\Psi^{-1})_{\eta\nu}
        (U_{i})_{\eta\eta}
        \\
        &=
        \chiPattern{i}{\mu}
        \sum_{\nu = 1}^{P}
        (\Psi^{-1})_{\nu\mu}
        \chiConjugatePattern{i}{\nu}\,\,\, .
    \end{align}
\end{subequations}
Therefore, if we now substitute the expansion in the $J_{ij}$ matrix expressed in terms of the effective pattern direction, we obtain

\begin{equation}\label{eq:expansion_coupling_matrix}
    J_{ij} = \dfrac{1}{N} \sum_{\mu = 1}^{P} c_{i}^{(\mu)} \xi_{ij}^{(\mu)} = \dfrac{1}{N} \sum_{\mu = 1}^{P} c_{i}^{(\mu)} \chiConjugatePattern{i}{\mu} \chiPattern{j}{\mu} \,\,\, .
\end{equation}

Substituting the solution of the coefficients $c_{i}^{(\mu)}$ we find the following expression for the coupling matrix
\begin{equation}
    J_{ij} = \dfrac{1}{N} \sum_{\mu,\nu}\chiPattern{j}{\mu}  (\Psi^{-1})_{\nu\mu} \chiConjugatePattern{i}{\nu}  \,\,\, ,
\end{equation}
where we have substituted the coefficients $c_{i}^{(\mu)}$ in the expansion of the coupling matrix \eqref{eq:expansion_coupling_matrix}.

To make the result coincide with the definition presented in Equation \eqref{def:pseudoinverse_complex} we swap the indices $\mu$ and $\nu$ to get
\begin{equation}
     J_{ij} = \dfrac{1}{N} \sum_{\mu,\nu} \chiConjugatePattern{i}{\mu} (\Psi^{-1})_{\mu\nu} \chiPattern{j}{\nu} \,\,\, .
\end{equation}
It is important to note that $J_{ij}$ is Hermitian (shown in in Appendix \ref{sec:appendix_b_properties}) and coincides with this notation the original definition in Equation \eqref{def:pseudoinverse_complex}.

To conclude the proof we show that the quadratic form in Equation \eqref{eq:quadratic_form_final} holds, from the quantities we have introduced to show the reconstruction process.
Given the rescaled coefficients $\cTilde_{i}^{(\mu)}$, we have for the pattern vector $\mathbf{\cTilde}_{i} = (c_{i}^{(1)}, \dots, c_{i}^{(P)})$ for the $i$-th oscillator
\begin{equation}\label{eq:quadratic_form_final}
    \mathbf{\cTilde}_{i}^{\dagger} \Gamma_{i} \mathbf{\cTilde}_{i} = 
    \mathbf{\cTilde}_{i}^{\dagger} U_{i}^{\dagger} \Psi^{*} U_{i} \mathbf{\cTilde}_{i} = 
    (U_{i} \mathbf{\cTilde}_{i})^{\dagger} \Psi^{*} U_{i} \mathbf{\cTilde}_{i} > 0 \,\,\, ,
\end{equation}
which holds because since the matrix $U_{i}$ is unitary, then $\Gamma_{i}^{\mu \nu}$ is Hermitian when $\Psi_{\mu\nu}$ is Hermitian, which is the first of two properties that we show in Appendix \ref{sec:appendix_b_properties}.
Since $\Psi_{\mu\nu}$ is Hermitian by construction and positive definiteness is guaranteed by the linear independence of the patterns, then $U_{i}^{\mu\nu} c_{i}^{(\nu)} \neq 0$ for all non-zero coefficient vectors $\boldsymbol{\cTilde}_{i} = \{\cTilde_{i}^{(\mu)}\}_{\mu = 1}^{P}$.
This concludes the proof as it is the exact equivalent quadratic form result of the proof by Diederich and Opper \cite{diederich_learning_1987}.
\subsection{Capacities}\label{ssec:04c_capacities}
We quantify the \textit{capacity} $C$ as the ratio between the stored patterns $P$ and number of oscillators $N$ and is in principle a function of both
\begin{equation}\label{def:capacity}
    C(N, P) := \dfrac{P}{N} \,\,\, .
\end{equation}
The \textit{retrieval capacity} of the system coincides with the \textit{storage capacity}, when the patterns are perfectly uncorrelated.
The reason for choosing the Pseudoinverse as a learning rule is precisely due to the property of decoupling patterns with respect to one another.

\subsection{Dynamics of Pattern Retrieval with the Kuramoto model}\label{ssec:04d_dynamics_kuramoto}
The dynamics that we use to perform the retrieval of stored patterns are simulated using the Kuramoto model \cite{kuramoto_formation_1975}.
The system of oscillators starts out from an initial condition where a gaussian mask with standard deviation $\sigma$ has been applied to an initial pattern as $\varphi_{i}^{(\mu)}(0) = \alpha_{i}^{(\mu)} + \mathcal{N}(0, \sigma)$.
The phases evolve towards a target pattern as
\begin{equation}\label{eq:kuramoto_model_pattern}
    \dot{\varphi}_{i}^{(\mu)} = \omega^{(0, \mu)}_{i} - \sum_{j = 1}^{N} \Imag \bigg\{ J_{ij} e^{\ii(\varphi_{i}^{(\mu)} - \varphi_{j}^{(\mu)})}  \bigg\} \,\,\,,
\end{equation}
where we assume the natural frequencies $\omega_{i}^{(0, \mu)}$ and for all patterns $\mu \in \{1, \dots, P\}$ and all oscillators $i$ to be the same, without loss of generality.
This way they can be considered an integration constant and thus neglected.
Since the evolution of the Kuramoto model is one of a set of coupled ordinary differential equations (ODEs), we have  that the dynamics are comparable to the one of parallel updates in spin-based models \cite{kanter_associative_1987}.

Using this setup we perform simulations in Section \ref{sec:05_simulations} on real and complex patterns, using different learning rules.
For the final state we measure the synchronization overlap defined in Equation \eqref{def:normalized_synchronization_overlap_trace} as an order parameter to assess how close the system steady-state has converged to the intended pattern.

\section{Simulation results}\label{sec:05_simulations}
We show the results for two kinds of sets of patterns for retrieval using the Kuramoto model.
This ODE evolves synchronously and in a parallel manner, conversely to Ising spin-glass models, where the behavior can also take place asynchronously with sequential dynamics \cite{kanter_associative_1987}.

The real-valued data is simulated using the matrix $J_{ij}$ shown in Equation \eqref{eq:kuramoto_model_real} in Section \ref{sec:03_background}, whereas the retrieval of complex-valued data is done using the form presented in Equation \eqref{eq:kuramoto_model_pattern}.
First, we compare the behavior of five learning rules on different sizes of the english alphabet datasets for the real case.
These simulations we performed are on custom created synthetic alphabet datasets.
The goal is to show that it is indeed possible to learn reliably with stable points grayscale patterns and retrieve them with Kuramoto oscillator dynamics.

Then, we show the results of the two best rules from a retrieval perspective, namely with $R \rightarrow 1$, for the extension to the complex case, where the oscillators are able to retrieve multi-state patterns using the Kuramoto model. 

To quantify the retrieval at steady-state indicated with $t^{*} = \infty$, we measure the synchronization overlap between $\varphi_{i}^{(\mu)}(\infty)$ and target pattern $\alpha_{i}^{(\nu)}$
using the definitions from Equations \eqref{def:complex_overlap_order_parameter} and \eqref{def:synchronization_overlap_order_parameter} to get the following metric
\begin{equation}\label{eq:steady_state_synchronization}
    \rho_{\mu\nu}^{\infty} = \dfrac{1}{N} \Bigg| \sum_{i = 1}^{N} e^{\ii  \varphi_{i}^{(\mu)}(\infty)} e^{-\ii \alpha_{i}^{(\nu)} } \Bigg| \,\,\, , \forall \mu , \nu .
\end{equation}
The metric we use to quantify and interpret the results is the normalized trace introduced in Equation \eqref{eq:steady_state_synchronization}.
Differently from the proof from Section \ref{sec:04_methods}, we set the diagonal $J_{ii} = 0$ for rule II, both with real and complex patterns as commonly done for parallel updates in binary spin-glass models \cite{kanter_associative_1987}.

The initial conditions for both the real and complex case are the same, we apply to the initial phase of each oscillator a value extracted from a Gaussian distribution with $\mathcal{N}(0, \sigma)$. Each steady-state value of the synchronization trace $R(\sigma)$ computed using definition \eqref{def:normalized_synchronization_overlap_trace} and Equation \eqref{eq:steady_state_synchronization}corresponds to a different value of $\sigma$.
We choose this approach for the initial conditions, since the dynamics of the Kuramoto model are continuous, conversely to the ones used by Hopfield \cite{Hopfield_1982} where state changes are discrete flips.

\subsection{Real learning rules for structured patterns}\label{ssec:05a_real_learning}
\begin{figure}[htbp]
    \centering
    \includegraphics[width = \linewidth]{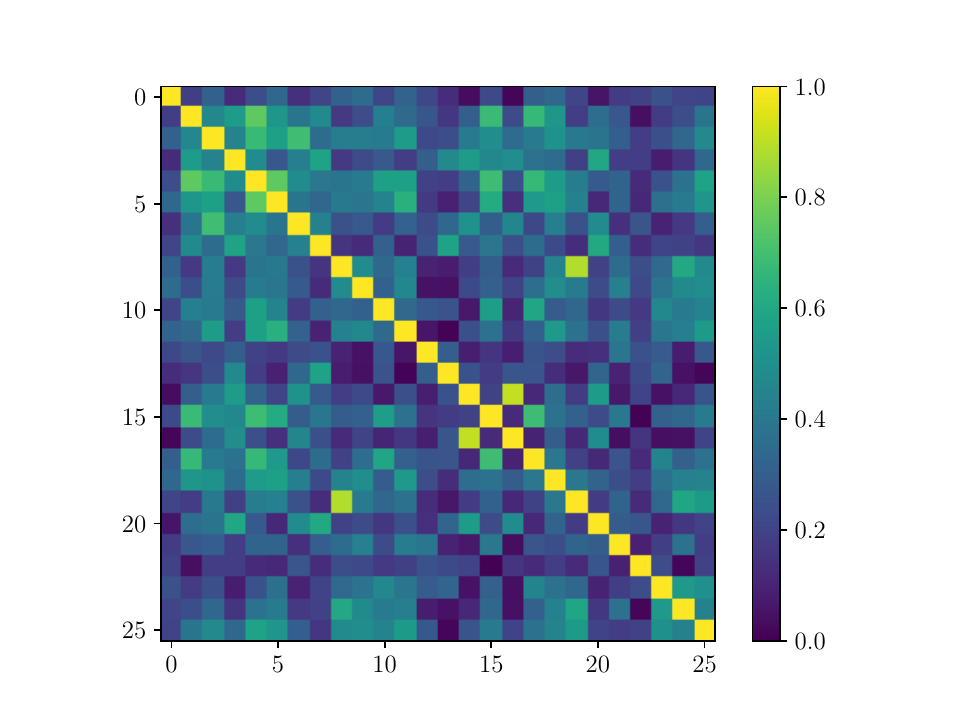}
    \caption{The \textit{synchronization overlap} $\rho_{\mu\nu}$ for the $22 \times 22$ pixel binary alphabet dataset (composed by $26$ patterns, one for each letter of the English alphabet) is represented as a heatmap which represents how similar patterns are to one another.
    The closer they are to one (e.g. the diagonal is the pattern with itself), the more similar, whereas the more dissimilar patterns are, the more the off-diagonal values of $\rho_{\mu\nu}$ will be close to zero.}\label{fig:dataset_rho_overlap_22x22}
\end{figure}

In order to find a good learning rule for patterns that have non-zero correlations, like structured datasets tend to have, we compare five learning rules that we label Hebbian \cite{1949_Hebb_organization_behavior, Hopfield_1982},
Storkey \cite{goos_increasing_1997}, Diederich and Opper rules I and II \cite{diederich_learning_1987}, and Pseudoinverse \cite{kanter_associative_1987}.
We compare the learning rules for three different dataset sizes, namely $N = \{ 42, 100, 484\}$.
Since the number of patterns of the alphabet we use is fixed, the capacity $C$ depends only on the number of oscillators of the network $N$.
It is important to remark that the similarity between the patterns can be quantified through the synchronization overlap \eqref{def:synchronization_overlap_order_parameter}.
An example for the real-valued case can be seen in Figure \ref{fig:dataset_rho_overlap_22x22}.

\begin{figure}[htbp]
    \centering
    \includegraphics[width = \linewidth]{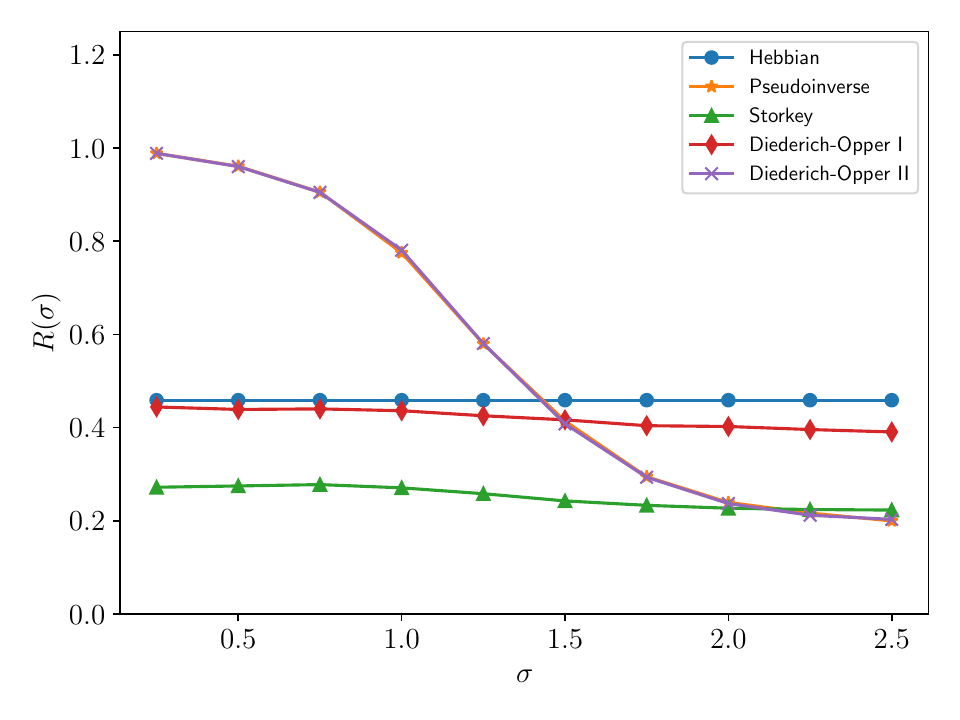}
    \caption{The simulations of the $7 \times 6$ network, where the capacity is $C \simeq 0.62$.}
    \label{fig:trace_overlap_function_noise_7x6}
\end{figure}

\begin{figure}[htbp]
    \centering
    \includegraphics[width = \linewidth]{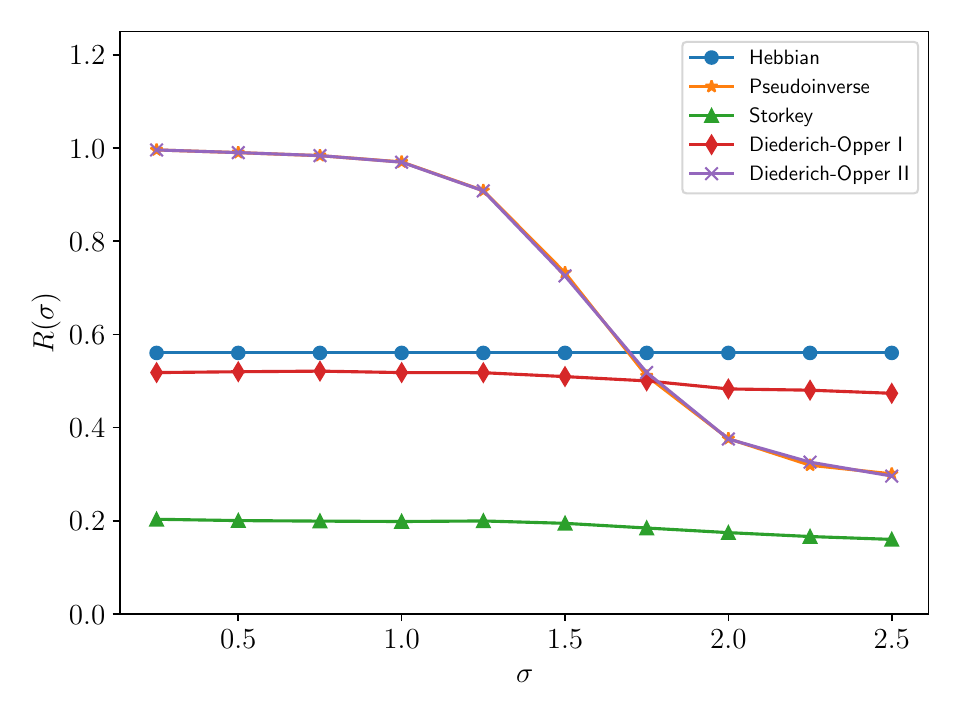}
    \caption{The simulations of the (this is quarter capacity) $10 \times 10$ network, where essentially at fixed $N$, the capacity $C \simeq \frac{1}{4}$.}
    \label{fig:trace_overlap_function_noise_10x10}
\end{figure}

\begin{figure}[htbp]
    \centering
    \includegraphics[width = \linewidth]{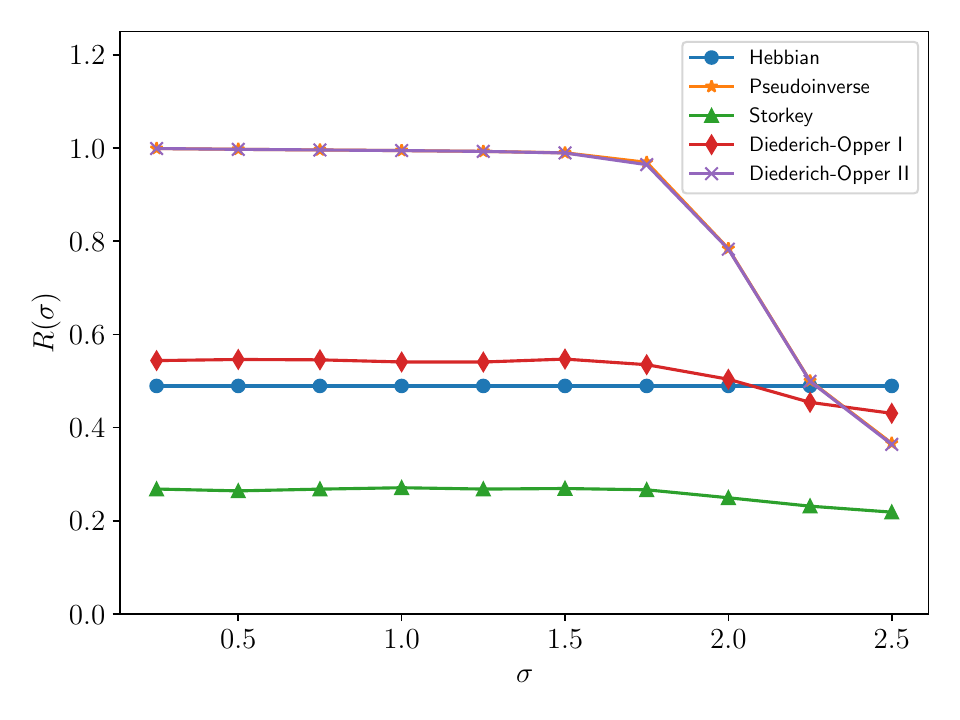}
    \caption{The simulations of the $22 \times 22$ network, with a capacity of $C \simeq 0.05$.}
    \label{fig:trace_overlap_function_noise_22x22}
\end{figure}

In order to quantify the retrieval quality of patterns through Kuramoto dynamics, we compute the trace of the synchronization $R(\sigma)$ between the steady-state and the memorized patterns as per Equation \eqref{eq:steady_state_synchronization}.
%
% RULE - HEBBIAN
It is possible to observe, how for all investigated network sizes in Figures \ref{fig:trace_overlap_function_noise_7x6}, \ref{fig:trace_overlap_function_noise_10x10}, and \ref{fig:trace_overlap_function_noise_22x22} the retrieval capacity of patterns of the Kuramoto model using the Hebbian learning rule \cite{1949_Hebb_organization_behavior} is constant. 
This can be explained by the fact that the weight matrix is "overloaded" or saturated, i.e. too many patterns are memorized, in addition to the patterns interfering, due to being correlated with each other.
The theoretical capacity of the Hopfield model with Hebbian learning \cite{Hopfield_1982} is $C(N) = \frac{N}{2\log(N)} \simeq 0.14 N$ in the thermodynamic limit obtained for random uncorrelated patterns in the case of a binary neural network \cite{amit_storing_1985}. 
As such it is not guaranteed one will be able to reliably retrieve the number of patterns they are memorizing, even if they are below the theoretical threshold.
This observation is even more true in the case of dynamics that differ from the ones used in the classical setting presented by Hopfield \cite{Hopfield_1982}.
What happens in this case at all sizes is that the initial conditions always flow to the same pattern, which is a superposition of all other memorized patterns, regardless of initial conditions.
Therefore, if one wishes to retrieve patterns from a correlated dataset using Kuramoto dynamics \eqref{eq:kuramoto_model_pattern} the Hebbian rule \cite{1949_Hebb_organization_behavior} is not a suitable construction for the weights.

% RULE - STORKEY
A modification of the Hebbian learning rule is the one introduced by Storkey \cite{goos_increasing_1997}.
This rule aims at improving the retrieval capacity, but from the the three dataset sizes performs poorly, with a value of $R(\sigma)$ that fluctuates around $0.2$.
This indicates that the learning rule is also not a suitable construction for retrieval with the Kuramoto model, despite an improved capacity of $C(N) \simeq \frac{N}{\sqrt{2 \log(N)}}$ in the large $N$ limit for random uncorrelated binary patterns on a Hopfield network \cite{goos_increasing_1997}.

% RULE - DIEDERICH-OPPER RULE I
The curve in Figures \ref{fig:trace_overlap_function_noise_7x6}, \ref{fig:trace_overlap_function_noise_10x10}, and \ref{fig:trace_overlap_function_noise_22x22} corresponding to the label Diederich-Opper I is the first of the two rules presented for binary spins in \cite{diederich_learning_1987}.
One can observe how the curve shifts upwards as a function of network size, capacity by extension, and the performance slightly worsens as a function of $\sigma$ for all sizes.
For no value of $N$ the curve reaches full synchronization, since even the largest value of $R(\sigma)$ lies below $0.6$, which is sufficient to observe and conclude that this learning rule is not a good construction for retrieval with Kuramoto dynamics.
A relevant remark by Diederich and Opper \cite{diederich_learning_1987} in a footnote is that rule I produces asymmetric networks, which seems to be incompatible with the continuous symmetry of the Kuramoto model.
Despite this, the rule may still work well in binary networks.

% RULE - PSEUDOINVERSE/DIEDERICH-OPPER RULE II
The Pseudoinverse learning rule \cite{kanter_associative_1987} aims at maximizing storage capacity such that $C(N) = 1$.
This can be achieved by decoupling the memorized patterns through the use of the concept of Pseudoinverse matrix.
This learning rule has the drawback of being neither local nor incremental, but this problem can be circumvented by recognizing that Diederich and Opper \cite{diederich_learning_1987} defined an incremental and local version of this learning rule for binary spin-glass networks.
Additionally, as shown in Section \ref{ssec:04b_analytical} this can be extended to complex unitary patterns, through the exploitation of the fact that the notion of Pseudoinverse is defined on a generic field $\mathbb{K}$ and therefore maintains its properties when taking $\mathbb{K} = \mathbb{C}$ \cite{penrose_generalized_1955}.
As a consequence, we show numerically that the retrieval capacities in Figures \ref{fig:trace_overlap_function_noise_7x6}, \ref{fig:trace_overlap_function_noise_10x10}, and \ref{fig:trace_overlap_function_noise_22x22} are so similar that the curves are superimposed.
Moreover, in terms of retrieval quality the purple and orange curves are the only ones changing as a function of the memorized patterns.
As expected for a curve of the synchronization trace $R(\sigma)$ that is close to $1$ for more values of $\sigma$, the more robust the retrieval capacity of the network is with respect to the initially applied mask.
From the obtained results for the retrieval performed with different learning rules, we conclude that the one that provides the best quantitative results of $R$ with Kuramoto dynamics are the Pseudoinverse \cite{kanter_associative_1987} and its local incremental version by Diederich and Opper \cite{diederich_learning_1987}.
This rule is robust and can reliably retrieve patterns with non-zero correlations through Kuramoto dynamics as shown in Section \ref{ssec:04d_dynamics_kuramoto}.
Therefore, we will focus all subsequent analysis on the behavior of these two rules, where the formulation by Diederich and Opper at convergence will reproduce the Pseudoinverse construction of the weight matrix as defined in Equations \eqref{def:pseudoinverse_complex} and \eqref{def:pseudoinverse_complex_alt}.
\subsection{Complex learning rules}\label{ssec:05b_complex_learning}
As shown in Section \ref{ssec:05a_real_learning}, the Kuramoto model \eqref{eq:kuramoto_model_pattern} performs best in the case of Pseudoinverse with a zeroed out diagonal \cite{kanter_associative_1987}.
The same result holds for the construction using the, namely rule II \cite{diederich_learning_1987}.
In order for the model of Equation \eqref{eq:kuramoto_model_pattern} to retrieve the patterns, it is important that information of patterns be encoded both in the strength of the couplings $|J_{ij}|$ and the phases $\arg\left(J_{ij}\right)$.
\subsubsection{Structured patterns}\label{sssec:05b_1_structured_patterns}
We constructed the dataset by using an alphabet from a font, which was then sampled by defining a size for the image one wants to use $N_{x}$ and $N_{y}$ to then obtain the total oscillator network size $N = N_{x} N_{y}$.
Then we apply a gaussian mask with a certain width $\sigma_{\text{mask}}$ to every element of the dataset normalize it to the range $[0, 1]$ and as shown in Section \ref{ssec:04a_mapping} multiply by $\pi$ to obtain the correct phase mapping using Equation \eqref{def:pixel_phase_mapping}.
All patterns are then stable up to a global rotation $\vartheta$ as pointed out in Section \ref{ssec:04a_mapping}.

\begin{figure}[htbp]
    \centering
    \begin{minipage}[b]{0.5\textwidth}
        \centering
        \includegraphics[width = \textwidth]{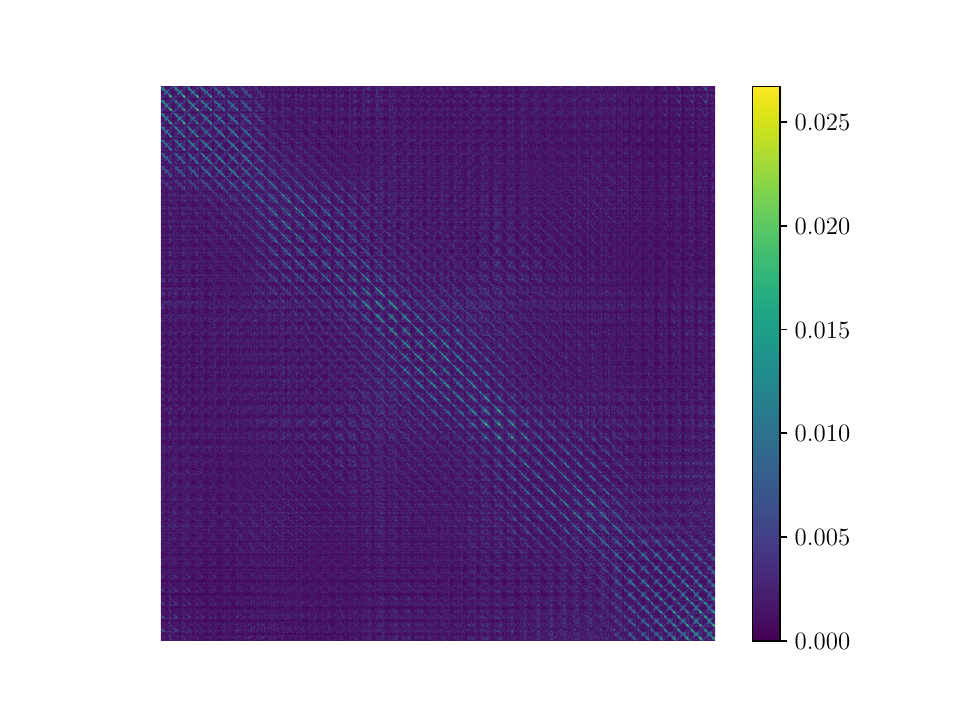}%\subcaption{(a)} \label{fig:weight_matrix_abs_42x42}
        \\(a) The absolute value $|J_{ij}|$ of the matrix $J_{ij}^{(\text{Pseudo})}$.
    \end{minipage}%
    \hfill 
    \begin{minipage}[b]{0.5\textwidth}
        \centering
        \includegraphics[width = \textwidth]{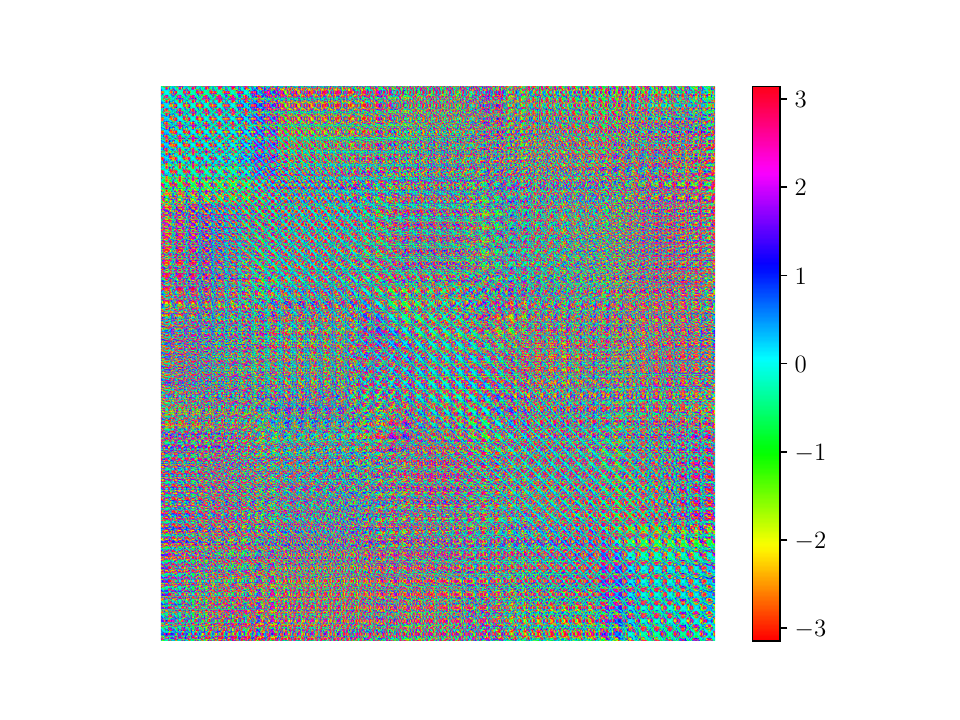} %\label{fig:weight_matrix_theta_42x42}
        \\(b) The angle part $\arg(J_{ij})$ of the matrix $J_{ij}^{(\text{Pseudo})}$.
    \end{minipage}
    \caption{ 
        This is the representation of the (a) absolute value $|J_{ij}|$ and (b) angle part $\arg(J_{ij})$ of the Pseudoinverse weight matrix for the $42 \times 42$ grayscale alphabet dataset.
        The values in (b) are represented using a circular color map due to periodicity of the phase values.
        In this example, the capacity of the weight matrix is in the low-storage regime, with $C(N) \simeq 0.015 \ll 1$.
    }\label{fig:weight_matrix_complex_42x42}
\end{figure}

\begin{figure}[htbp]
    \centering
    \begin{minipage}[b]{0.5\textwidth}%\columnwidth}
        \centering
        \includegraphics[width = \textwidth]{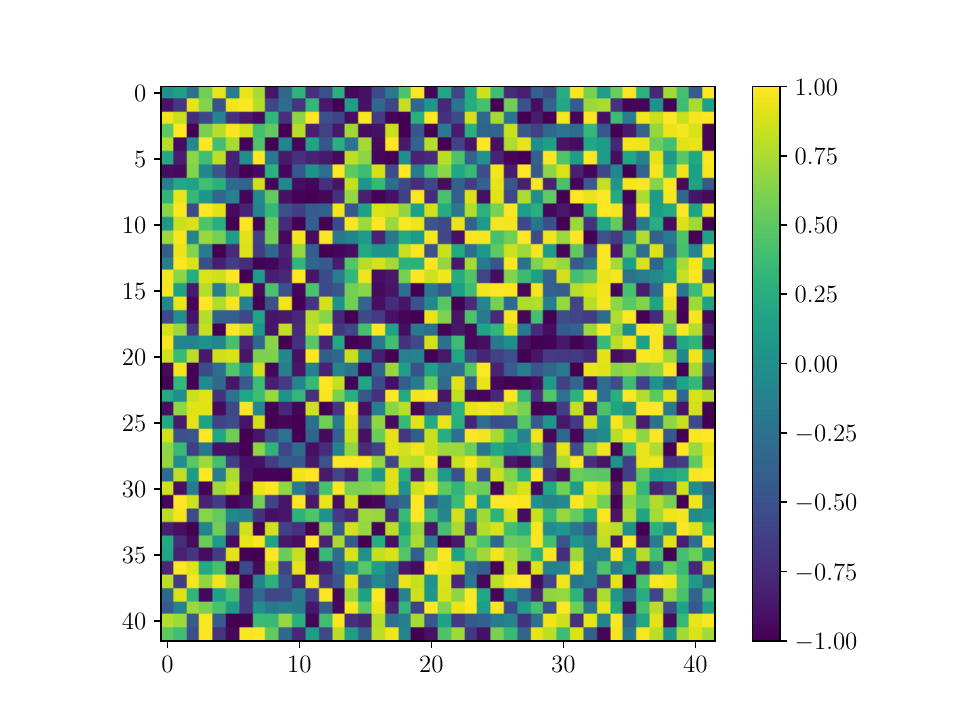}%\subcaption{(a)}
        % \caption{(a) Initial condition for the target pattern.}
        % No \caption command here
        \\(a) Example of initial condition of a target.
    \end{minipage}%
    \hfill 
    \begin{minipage}[b]{0.5\textwidth}
        \centering
        \includegraphics[width = \textwidth]{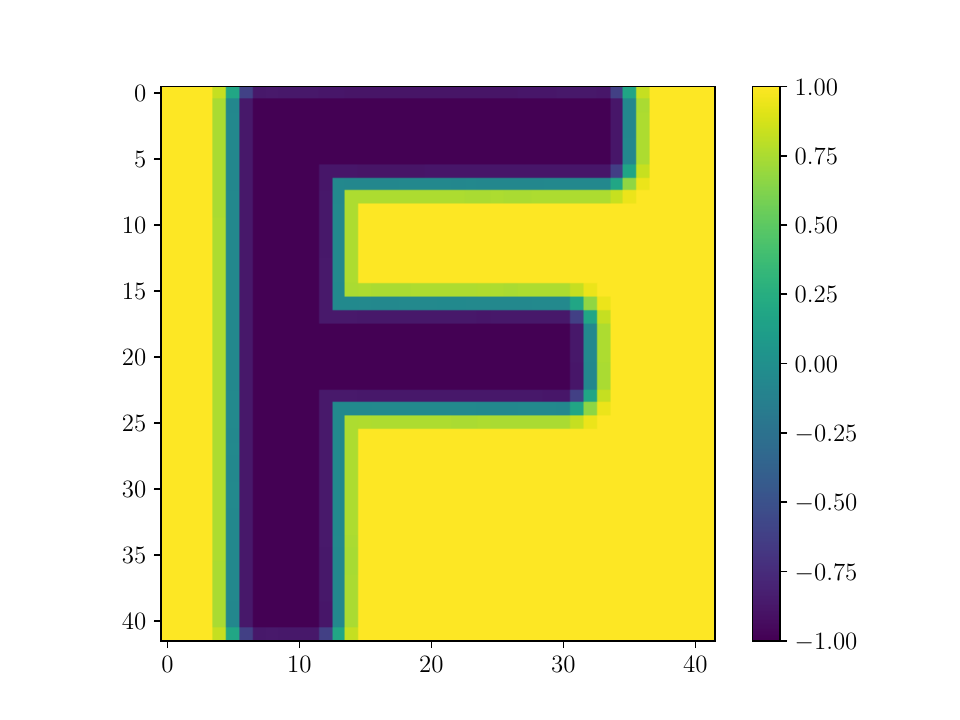}
        \\(b) Example of steady-state of the target pattern.
    \end{minipage}
    \caption{ 
        Initial condition (a) and Steady state (b) of the letter F in the alphabet dataset for a $N = 42 \times 42$ oscillator network.
        Figure \ref{fig:pattern_retrieved} (a) shows a Gaussian mask with a standard deviation of $\sigma = 1.75$ applied to the target pattern to generate an initial condition.
        Figure \ref{fig:pattern_retrieved} (b) shows the retrieved pattern coinciding with the target pattern.
        It is possible to observe how the retrieved stable pattern displays a gradient around the boundary with values in the interval $[-1, 1]$.
    }\label{fig:pattern_retrieved}
\end{figure}

As established in Section \ref{ssec:05a_real_learning} the best retrieval results are found with a weight matrix constructed using Equation \eqref{def:pseudoinverse_complex}.
Therefore, we focus further investigations on the weight matrix obtained through the Pseudoinverse and the method from Section \ref{ssec:04b_analytical}.
The reason why the former method is relevant, is that for biological plausibility one requires new connections to be created through local and incremental steps.
In Figures \ref{fig:weight_matrix_complex_42x42} (a) and (b), one can see the modulus and angle parts of the complex weight matrix used to store the patterns into $J_{ij}$.
Additionally, an example of a retrieved pattern, starting from the initial conditions from Figure \ref{fig:pattern_retrieved} (a), can be seen in Figure \ref{fig:pattern_retrieved} (b) for the grayscale dataset of size $N = 42 \times 42$.

\begin{figure}[htbp]
    \centering
    \begin{minipage}[b]{0.5\textwidth}
        \centering
        \includegraphics[width = \textwidth]{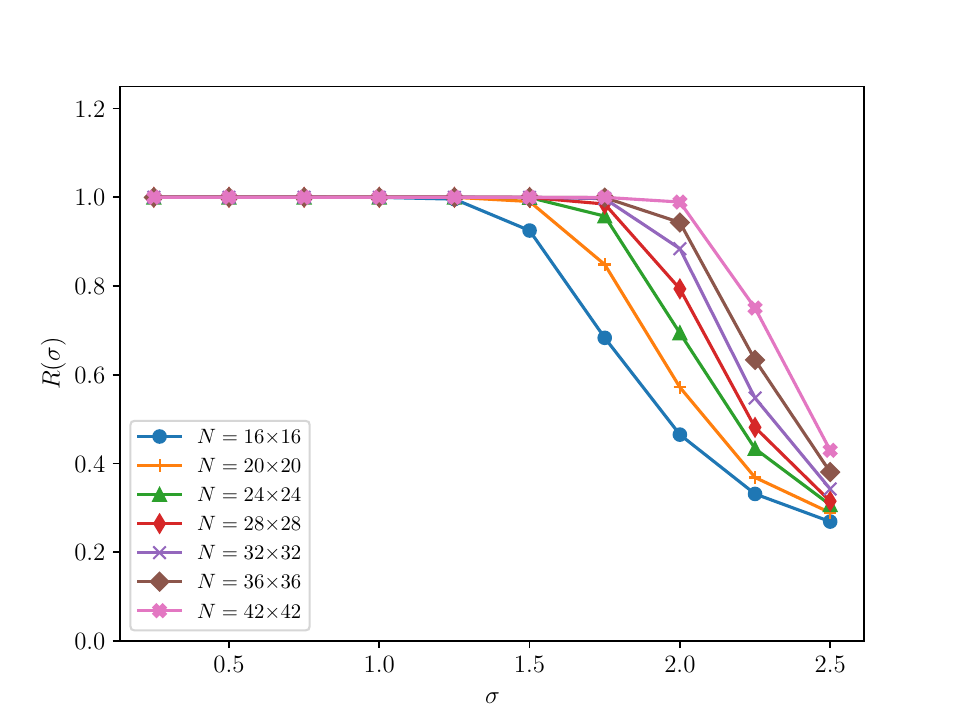}%\subcaption{(a)}
        \\(a) Synchronization trace $R(\sigma)$ for \textit{complex Pseudoinverse} as a function of initial spread of the Gaussian mask $\sigma$.
    \end{minipage}%
    \hfill 
    \begin{minipage}[b]{0.5\textwidth}
        \centering
        \includegraphics[width=\textwidth]{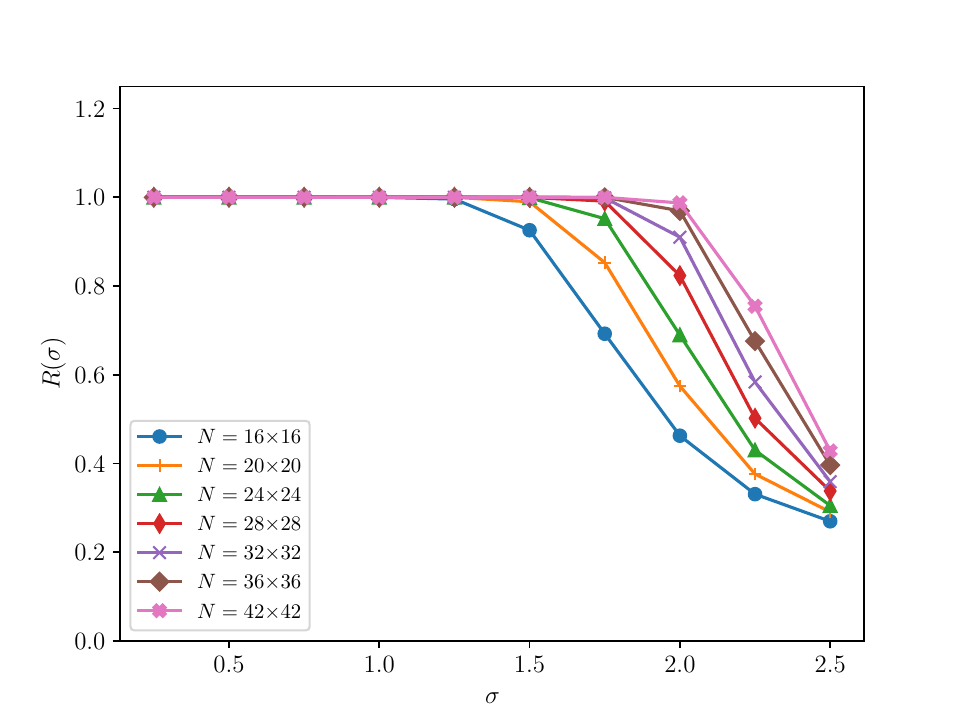}
        \\(b) Synchronization trace $R(\sigma)$ for \textit{complex rule II} as a function of initial spread of the Gaussian mask $\sigma$.
    \end{minipage}
    \caption{ 
        The simulations of the (a) \textit{Pseudoinverse} and (b) \textit{complex rule II} weight matrices encoding an alphabet dataset for different sizes were all performed for $10$ oscillation cycles $n_{c}$ for all sizes and $1000$ time steps $n_{t}$, except for the $42 \times 42$ dataset which was set to $2000$ time steps and $5$ oscillation cycles.
        One period is defined as the full period of a sine function and it is $2 \pi$.
        Overall the number of numerical points on the time-stencil is the same and equates to $n_{c} n_{t} = 10000$.
        The number of random initial conditions is determined by the standard deviation of the Gaussian mask $\sigma$.
        We then test each pattern of the set on is $100$ for each data point.}\label{fig:trace_overlap_function_noise_pseudoinverse_rule_ii_complex}
\end{figure}

From the plots of Figures \ref{fig:trace_overlap_function_noise_pseudoinverse_rule_ii_complex} (a) and (b) one can see the retrieval quantified through Equation \eqref{def:normalized_synchronization_overlap_trace} yield numerically comparable results.

Every curve represents a different network size, namely a value of capacity as a function of $N$ alone since the patterns in the datasets is fixed.
One can observe how for increasing $N$ the retrieval is more robust with respect to the initially applied mask to the target pattern.
For the $16 \times 16$ network $R$ decreases around a value of $\sigma = 1.25$, and the largest simulated network $42 \times 42$ starts decreasing at $\sigma = 2.0$.
This is consistent with the expected behavior of retrieval, since at higher number of oscillators in the network the ratio between stored patterns $P$ and number of oscillators $N$ is lower.
As a consequence the same amount of patterns have more redundancy in the larger sized weight matrix $J_{ij}$.
In the case of \textit{rule II}, $R$ is close to $1$ for more corresponding values of $\sigma$.
This behavior might be due to having used a diagonal conditioning factor $\lambda = 10^{-6}$ for constructing the weight matrix with the method of Section \ref{ssec:04b_analytical}.
\begin{figure}[htbp]
    \centering
    \includegraphics[width = \linewidth]{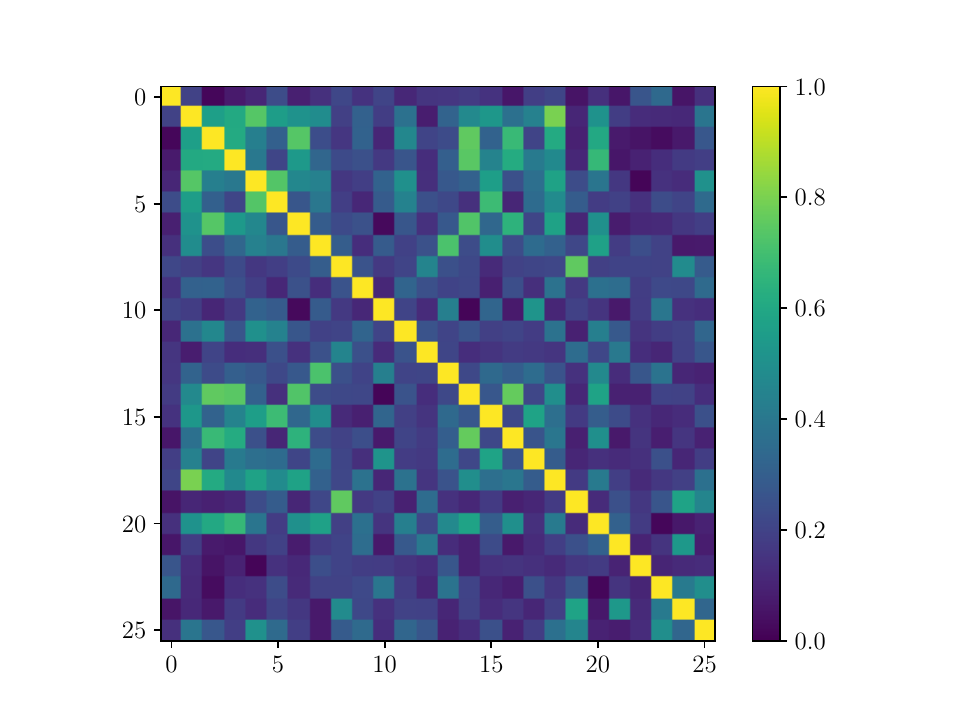}
    \caption{The \textit{synchronization overlap} matrix $\rho_{\mu\nu}$ for the $42 \times 42$ grayscale alphabet dataset is represented as a heatmap. 
    The closer the  values are to one (yellow) the more similar they are.
    Conversely, the more dissimilar patterns are the closer they will be to zero (dark blue).}
    \label{fig:dataset_rho_overlap_42x42}
\end{figure}
%
% With figure* it takes up both of the double columns

\subsubsection{Random patterns}\label{sssec:05b_2_random_patterns}
\begin{figure}[htbp]
    \centering
\includegraphics[width = \linewidth]{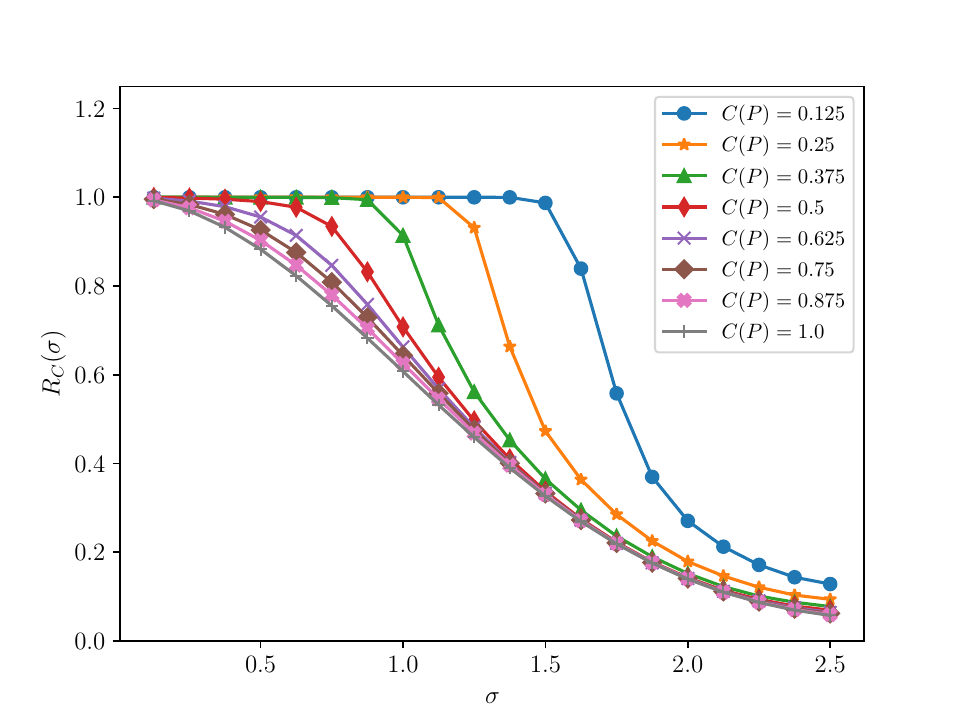}
    \caption{We plot the synchronization curves $R_{C}$ for a network of size $N = 484$ at different loading of memorized patterns using the matrix with complex patterns learned through the Pseudoinverse rule Equation \eqref{def:pseudoinverse_complex}.
    The index $C$ indicates that each curve is associated to a specific value of capacity, namely to a fraction of memorized patterns in the interval $\{0.125, 0.25,\dots, 1.0\}$.
    Each data-point corresponds to $n = 100$ runs.
    We neglected the point at $\sigma = 0$, since the patterns themselves are stable by construction, when no noise is applied.}\label{fig:trace_overlap_random_22x22_function_noise_alpha_loads}
\end{figure}

Before discussing the setup and simulation results, we define the random complex dataset first.
We generate the random patterns by subdividing the semicircle as shown in Figure \ref{fig:unit-semicircle-pixels-to-phases} and Equation \eqref{def:pixel_phase_mapping} between $0$ and $\pi$ into $n_{L} = 8$ grayscale levels.
The values are extracted according to a uniform distribution $\alpha_{i}^{(\mu)} \sim \mathcal{U}([0, \pi])$, which then are mapped using Equation \eqref{eq:mapping_phases_to_complex_patterns} to complex patterns.
Then we tested the retrieval dynamics of the generated random complex patterns defined according to equation \eqref{def:complex_spin_pattern} using the Kuramoto model \eqref{eq:kuramoto_model_pattern}.
For the random dataset simulations we choose to keep the systems size $N$ fixed to $484$ oscillators.
Consequently, we evaluate the retrieval performance through the synchronization trace $R$ for different values of capacity.
Therefore, $C(P)$ is a function of the number of memorized patterns $P$ alone.
We chose this to compare the performance with that of the largest binary network size for which the real and complex synthetic alphabet datasets could be simulated.
In Figure \ref{fig:trace_overlap_random_22x22_function_noise_alpha_loads} one can observe the synchronization trace behavior for eight different values of capacity $C$, with increasing number in stored patterns $P$ inside the weight matrix.
The synchronization curves for the network of size $N = 484$ present $n = 100$ randomizations of the initial condition per run.
We neglected the point at $\sigma = 0$, since the patterns themselves are stable by construction, when no noise is applied.
In the case $C(P) \leq 0.5$, it is possible to observe the synchronization curves $R_{C}$ of Figure \ref{fig:trace_overlap_random_22x22_function_noise_alpha_loads} are more distant than for $C > 0.5$, where they cumulate towards the curve for $C(P) = 1$.
Moreover, when the memorized patterns exceed half capacity $C(P) > 0.5$, the curves are less resilient to initial conditions. 
Thus, it becomes less likely to have synchronization to target patterns for finite $\sigma$.
As a result, we find that numerically there is always a finite region of values for $\sigma$, that displays correct retrieval, which coincides with synchronization to the target $R_{C} = 1$, and a vanishing behavior for $C(P) \rightarrow 1$.
Therefore, the observed behavior is consistent with what is expected from pattern retrieval dynamics at zero temperature, where the region where $R_{C} = 1$ shrinks as $C(P)$ tends to $1$.

\section{Conclusions}\label{sec:06_conclusions}
In this paper, we have identified the best candidate for a learning rule to be used with Kuramoto dynamics.
We then tested on three different sized binary pattern sets the retrieval accuracy and quantified it through the synchronization trace $R$.

This observation led us to identify the Pseudoinverse \cite{kanter_associative_1987} as a candidate for the central result.
Consequently we extended the method to obtain a weight matrix $J_{ij}$ with unitary complex spin patterns using a local and incremental construction.
This is an extension of a previously known result \cite{diederich_learning_1987} from binary spins to complex patterns.

Moreover, we have shown how retrieval with the Kuramoto model is possible using the Pseudoinverse weight matrix construction from Equation \eqref{def:pseudoinverse_complex} for real- and complex-valued structured sets of patterns.
The behavior was analyzed for different values of $N$. 
For random patterns we quantified how the network behaved as a function of stored patterns $P$.
In the first case, retrieval improved for larger networks and in the second retrieval deteriorated with the number of stored patterns in the matrix.

As a consequence of the results we presented, some potential future outcomes will be the investigation of higher order networks \cite{krotov_dense_2016, demircigil_model_2017, 2025_berloff_higher_order_kuramoto_oscillator_network, delacour_lagrange_2025} combining the Kuramoto model with modern Hopfield networks.
The Pseudoinverse method we analyzed in this work offers a robust construction to eliminate cross-talk between patterns and increase the retrieval capacity.
These approaches can therefore be merged, to model higher order complex-valued higher order networks of Kuramoto oscillators.
\section{Acknowledgements}\label{sec:09_ackowledgements}
This work has received funding from the European Research Council ERC Consolidator Grant, THERMODON, with grant ID number 101125031.

\appendix
\section{Correspondence between symbols}\label{sec:appendix_a_translation}
To help bridge between the extension of the binary formulation by Diederich and Opper \cite{diederich_learning_1987} to our proof in Section \ref{ssec:04b_analytical}, we provide Table \ref{tab:symbol_translation}.
In this we relate the symbols used for equivalent quantities in the two works.
\begin{table}[htbp]
\centering
\begin{tabular}{lll}
\toprule
Physical quantity & Diederich-Opper \cite{diederich_learning_1987} & This work \\
\midrule
\textit{Spin pattern} & $S_{i}^{\mu} \in \mathbb{Z}_{2} $ & $\chiPattern{i}{\mu} \in U(1)$ \\
\textit{Effective pattern direction} & $\sigma_{j}^{\mu}$ & $\xi_{ij}^{(\mu)}$ \\
\textit{Embedding strengths} & $x^{\mu}$ & $c_{i}^{(\mu)}$ \\
\textit{Local field} & $E_{i}^{\mu}$ & $h_{i}^{(\mu)}$ \\
\textit{(Complex) Overlap} & $C_{\mu\nu}$ & $\Psi_{\mu\nu}$ \\ 
\textit{Reconstruction matrix} & $B^{\alpha\beta}$ & $\Gamma_{i}^{\mu\nu}$ \\
\textit{Weight matrix} with\\ $i$-th index fixed & $I_{j}$ & $J_{ij}$ \\
\bottomrule
\end{tabular}
\caption{Symbol translation between \cite{diederich_learning_1987} and this work.}
\label{tab:symbol_translation}
\end{table}
\\
%
% Properties
\section{Properties of complex overlap and weight matrix}\label{sec:appendix_b_properties}
We show the Hermiticity of the complex overlap $\Psi_{\mu\nu}$ and of the complex Pseudoinverse weight matrix $J_{ij}^{(\text{Pseudo})}$.

\begin{property}[$\Psi$ is Hermitian]\label{prop:psi_hermitian}
    The complex overlap matrix $\Psi_{\mu\nu}$ is Hermitian.
\end{property}
\begin{proof}
    The matrix $\Psi$ is Hermitian by construction, which we show by taking the conjugate transpose
    \begin{subequations}
        \begin{align}
            (\Psi_{\nu\mu})^{*} &= \left( \dfrac{1}{N}\sum_{k = 1}^{N} \chiPattern{k}{\nu} \chiConjugatePattern{k}{\mu}\right)^{*}
            \\
            &= \dfrac{1}{N}\sum_{k=1}^{N}\chiPattern{k}{\mu}\chiConjugatePattern{k}{\nu}
            \\
            &= \Psi_{\mu\nu} \,\,\, .
        \end{align}
    \end{subequations}
    Which implies $\Psi^{\dagger} = \Psi$ and concludes the proof.
\end{proof}

\begin{property}[$J_{ij}$ is Hermitian]
    The coupling matrix $J_{ij}$ is Hermitian.
\end{property}
\begin{proof}
    In the Hermitian Pseudoinverse construction, the coupling matrix is defined in Equation \eqref{def:pseudoinverse_complex} is itself Hermitian.
    We verify this by taking the complex conjugate and transposing the matrix itself from the definition
    \begin{subequations}
        \begin{align}
            (J_{ji})^{*} &=
            \left(\dfrac{1}{N}\sum_{\mu, \nu}^{P}
            \chiConjugatePattern{j}{\mu}(\Psi^{-1})_{\mu\nu}\chiPattern{i}{\nu}\right)^{*} 
            \\
            &=
            \dfrac{1}{N}\sum_{\mu, \nu}^{P}
            \chiPattern{j}{\mu}\left((\Psi^{-1})_{\mu\nu}\right)^{*}
            \chiConjugatePattern{i}{\nu} \,\,\, .
        \end{align}
    \end{subequations}
    Since $\Psi$ is Hermitian with $\left((\Psi^{-1})_{\mu\nu}\right)^{*} = \Psi^{-1}_{\nu\mu}$from Property \ref{prop:psi_hermitian}, its inverse $\Psi^{-1}$ is also Hermitian.
    This in turn implies that the following holds
    \begin{align}
        (J_{ji})^{*} &= \dfrac{1}{N}\sum_{\mu, \nu}^{P}
            \chiPattern{j}{\mu}(\Psi^{-1})_{\nu\mu}
            \chiConjugatePattern{i}{\nu} \,\,\, .
    \end{align}
    If we relabel the indices $\mu$ and $\nu$ we find the following
\begin{subequations}
        \begin{align}
            (J_{ji})^{*} &=
            \dfrac{1}{N}\sum_{\mu, \nu} \chiConjugatePattern{i}{\mu}(\Psi^{-1})_{\mu\nu}\chiPattern{j}{\nu} = J_{ij} \,\,\,
            \\
            &\implies \mathbf{J}^{\dagger} = \mathbf{J} \,\,\, ,
        \end{align}
\end{subequations}
    which concludes the proof that the coupling matrix $J_{ij}$ defined through the Pseudoinverse rule for unit-modulus complex patterns is Hermitian.
\end{proof}

\bibliographystyle{apsrev4-2}  % or aipauth for author-year
\bibliography{bibliography_sbravati}

\end{document}